\documentstyle[12pt,aps,prb,epsf]{revtex}
\input tcilatex

\QQQ{Language}{
American English
}

\begin{document}

\bigskip

\begin{center}
{\huge {\bf The First Acoustics Peak}}

\bigskip 

{\huge {\bf in CMBR and}}

\bigskip 

{\huge {\bf Cosmic Total Density}}

\bigskip \bigskip
\bigskip \bigskip 

{\Large {\bf De-Hai Zhang}}\\[0pt]

\bigskip 

(e-mail: dhzhang@sun.ihep.ac.cn)\\[0pt]

Department of Physics, Graduate School, Academia Sinica,\\[0pt]

P.O.Box 3908, Beijing 100039, P.R.China.\\[0pt]

\bigskip

{\bf Abstract:}
\end{center}

\baselineskip=16pt

{\hspace*{5mm}Boomerang measured the first peak in CMBR to be at location of}
$l_D=$ $196\pm 6$, which excites our strong interesting in it. A widely
cited formula is $l_D\simeq 200\Omega _T^{-0.50}$ to estimate the cosmic
total density. Weinberg shows it is not correct and should be $l_D\propto
\Omega _T^{-1.58}$ near the interest point $(\Omega _m,\Omega _\Lambda )$ $%
=(0.3,0.7)$. We show further that it should be $l_D\propto \Omega
_T^{-1.43}\Omega _m^{-0.147}$ or $\Omega _T^{-1.92}\Omega _\Lambda ^{0.343}$
near the same point in the more veracious sense if we consider the effect
from the sound horizon. We draw a contour graph for the peak location, show
that the recent data favor to a closed universe with about $\Omega _T\simeq
1.03$. If we insist on obtaining a flat universe, a point $(0.36,0.64)$,
i.e., more matter and less vacuum energy, is still possible, which has a
more right-side first peak $l_D=208$ in CMBR and a smaller acceleration
parameter $-q_0=0.10$ for the $z=0.4$ redshift SNIa.

\bigskip

\baselineskip=20pt

\section{Introduction}

{\hspace*{7mm}}Recently, the Boomerang$^{[1]}$ has observed that there exits
a vivid peak structure in the power spectrum of Cosmic Microwave Background
Radiation (CMBR) anisotropy. The first acoustics peak appears at location of
Legendre multipole $l_D=196\pm 6$, then they obtained their conclusion that
the universe is almost Euclidean flat, i.e., the total density is near
critical, $\Omega _T=\Omega _m+\Omega _\Lambda \simeq $ $1.00\pm 0.12$,
which is a most important cosmological parameter concerned by us. People
often uses a widely cited formula $l_D\simeq 200/\sqrt{\Omega _T}$ to
estimate the cosmic total density if one knows the first peak location$%
^{[2,3]}$. However, Weinberg shows that this formula is not even a crude
approximation in the greatest current interest region$^{[4]}$. He says that
this formula should be $l_D\propto \Omega _T^{-1.58}$ near the favorite
point $(\Omega _m,\Omega _\Lambda )$ $=(0.3,0.7)$. Weinberg's work start a
precedent of how to analysis the complicated phenomena of the acoustics
peaks in CMBR in a simple way. It is a significant for us how to hold the
physical essential by using as possible as fewer calculations. Then this
puts forward an important question, i.e., we must be careful to analysis the
relation between the first peak location and the cosmological parameters $p_i
$, such as the matter density $\Omega _m$ (which includes both of the baryon
density $\Omega _b$ and the cold dark matter density $\Omega _d$), radiation
density $\Omega _\gamma $, the vacuum energy density (e.g., cosmological
constant) $\Omega _\Lambda $, and the redshift $z_r$ of the recombination
epoch. We also concern of how much value is the proportional coefficient in
Weinberg's formula.

In this paper we shall consider a neighbor analysis based on
Efstathiou-Bond's formula to calculate the position of the first peak. This
analysis is more precious and will give a reappearance of the Weinberg's
phenomenon$^{[4]}$, i.e., the acoustics peak provides a more stringent 
constraint on $\Omega _T$ than an usual expected case.

\section{The position of the first peak}

The calculation of the first peak location is very complicated if we use the
relevant sets of the pertubative evolution equations in the cosmology$^{[5]}$%
. Efstathiou and Bond have obtained a good experiential formula for the
first peak location$^{[6]}$, we can rewrite it as the following in a more
clear way with pertinent variable dependent,%
$$
l_D(p_i)=C_{eb}\cdot A(p_i)\cdot |\Omega |_k^{-1/2}\text{Sinn}[|\Omega
|_k^{1/2}\int_1^{z_r}B(w,p_i)dw],\eqno{(1)} 
$$
where the function ''Sinn'', which origins from the angular diameter
distance to the last scattering surface, means that if $\Omega _k>0$, Sinn$%
[f]=\sinh [f]$; if $\Omega _k<0$, Sinn$[f]=\sin [f]$; if $\Omega
_k\rightarrow 0$, Sinn$[f]\rightarrow f$. Here $\Omega _k=1-\Omega _m-\Omega
_\Lambda $ is the curvature term density. The function $A(p_i)$
origins from the sound horizon of the photon-baryon fluid before the
recombination, 
$$
A(p_i)=\frac{3\pi }4\sqrt{\dfrac{\omega _b\Omega _m}{\omega _\gamma }}\cdot
\left( \ln \dfrac{\sqrt{\Omega _m(4\omega _\gamma z_r+3\omega _b)}+\sqrt{%
3\omega _b(\omega _\gamma h^{-2}z_r+\Omega _m)}}{\sqrt{\omega _\gamma z_r}%
\cdot (\sqrt{4\Omega _m}+\sqrt{3\omega _bh^{-2}})}\right) ^{-1},\eqno{(2)} 
$$
and the function $B(p_i)$ is relative with the cosmic comformal expansion
ratio,%
$$
B(w,p_i)=(\Omega _\Lambda +(1-\Omega _m-\Omega _\Lambda )w^2+\Omega
_mw^3+\omega _\gamma h^{-2}w^4)^{-1/2}.\eqno{(3)} 
$$
Hereafter we use only the parameters $\omega _\gamma =\Omega _\gamma h^2$
and $\omega _b=\Omega _bh^2$ rather than $\Omega _\gamma $ and $\Omega _b$
due to former higher accuracy. We know $\omega _\gamma =$ $4.31\times 10^5$
exactly$^{[7]}$ from the background temperature $2.73^oK$, and the error of $%
\Omega _\gamma $ comes mainly from the Hubble constant $100h$ km$\cdot $sec$%
^{-1}\cdot $Mpc$^{-1}$. We know $\omega _b$ more accurately than $\Omega _b$
from the Big Bang Nucleosynthesis. So that we must consider the $l_D$ error
originated from an uncertainty of the Hubble constant $h$, which appears in
Eqs.(2-3). An important coefficient $C_{eb}$ appears in the formula of $l_D$
expressed by us. I call it as the Efstathiou-Bond coefficient. Its
origination is the projection from the three-dimensional temperature power
spectrum to a two-dimensional angular power spectrum, its value is taken as $%
C_{eb0}=0.746$ by Efstathiou and Bond. In order to adapt to the need of the
future more accurate calculation or modification, we take it as $%
C_{eb}=c\cdot C_{eb0}$ and $c$ is a constant to be established. If $c$ is$1$
it is corresponding to the above Efstathiou-Bond value.

\section{Neighbor analysis}

Considering various achievements come from the different fields of the
Cosmology, we choose a favorite point for various cosmological parameters $%
p_i$, which is $\Omega _{m0}\simeq 0.3$, $\Omega _{\Lambda 0}\simeq 0.7$, $%
\omega $$_{b0}\simeq 0.02$, $h_0\simeq 0.7$, $z_{r0}\simeq 1100$. In the
neighbor of any points, the $l_D$ can be expressed approximately as%
$$
l_D=l_D(p_{i0})\cdot \prod (\frac{p_i}{p_{i0}})^{I_i},\eqno{(4)} 
$$
the power indexes can be calculated by%
$$
I_i=\frac{\partial l_D}{\partial p_i}|_{p_{i0}}\cdot \frac{p_{i0}}{%
l_D(p_{i0})}.\eqno{(5)} 
$$

In the neighbor of the favorite point, relation between the first peak
location and the cosmological parameters is, in according to a direct
numerical calculation of Eqs.(1-5) by Mathematica,%
$$
l_D=l_{D0}(\frac{C_{eb}}{C_{eb0}})(\frac{z_r}{z_{r0}})^{0.670}(\frac
h{h_0})^{-0.487}(\frac{\omega _b}{\omega _{b0}})^{0.059}(\frac{\Omega _m}{%
\Omega _{m0}})^{-0.576}(\frac{\Omega _\Lambda }{\Omega _{\Lambda 0}}%
)^{-1.004},\eqno{(6)} 
$$
where $l_{D0}=213$. If we want to express it in terms of the parameter $%
\Omega _T$, we have 
$$
l_D\propto (\frac{\Omega _T}{\Omega _{T0}})^{-1.43}(\frac{\Omega _m}{\Omega
_{m0}})^{-0.147}\propto (\frac{\Omega _T}{\Omega _{T0}})^{-1.92}(\frac{%
\Omega _\Lambda }{\Omega _{\Lambda 0}})^{0.343}.\eqno{(7)} 
$$
We see that in the point $(\Omega _m,\Omega _\Lambda )=(0.3,0.7)$, this
power index is different from Weinberg's result. The main reason is that the
function $A(p_i)$ is still dependent on $\Omega _m$. Anyhow, the Weinberg's
conclusion, i.e., a more stringent constraint on $\Omega _T$ will be
provided by the acoustics peak position, is correct, which can be seen from
a large index absolute value about $\Omega _T$. If we only want to change $%
\Omega _T$ alone, the first formula in Eq.(7) means to fix $\Omega _m$ and
to vary $\Omega _\Lambda $, the second formula means to fix $\Omega _\Lambda 
$ and to vary $\Omega _m$.

However, the index varies quite a bit for other points, and rather
complicated for all parameters. The most important parameters are $\Omega _m$%
, and $\Omega _\Lambda $, therefore we fix the three parameters $z_r$, $h$
and $\omega _b$ at first. We choose our greatest current interest region as $%
0.2<$$\Omega _m<0.4$ and $0.6<$$\Omega _\Lambda <0.8$. In this region we can
obtain a good fitting by a simple formulas, which accuracy is higher than $%
0.5$ percent (notice that out of this region the error immediately becomes
large!),%
$$
l_D=c[71.4+(1485-9275\Omega _m+30646\Omega _m^2-51076\Omega _m^3+33844\Omega
_m^4)(1-0.86\Omega _\Lambda )],\eqno{(8)} 
$$
As a comparison, the accuracy of Eq.(6) is only $1$ percent in a small
region of $0.25<$$\Omega _m<0.35$ and $0.65<$$\Omega _\Lambda <0.75$.
Another important observation constraint on the cosmological parameters
comes from the cosmic deceleration parameter $q_0$, which can be expressed$%
^{[8]}$ as $q_0=0.8\Omega _m-0.6\Omega _\Lambda =-0.2\pm 0.1$ for the SNIa
with redshift $z\simeq 0.4$. A contour graph about $l_D$ is drawn in the
figure.1.

\section{Weinberg's phenomenon in contour graph}

This figure has the following feature. Near the favorite point $P1$, the $%
l_{D\text{ }}$contour (a straight line in fact) is not parallel with the $%
\Omega _T$ contour, we can see a clear cross between the two lines near this
point. When $\Omega _T$ approach $1.07$ or $l_D$ approach less than $190c$,
the both lines begin to parallel. In this case the $l_D$ is alone dependent
on $\Omega _T$. Only in this case we can express $l_D\propto \Omega _T^{I_T}$%
, but it has not already been a flat universe. The point $P2$ is the cross
point of $\Omega _T=1$ and $q_0=-0.20$. The point $P3$ is a cross point of $%
l_D=200c$ and $q_0=-0.20$, which has value about $\Omega _T=1.03$, i.e., if
the formula (1) is correct (specially for Efstathiou-Bond coefficient) to
calculate the first peak position, then the recent result of the first peak
squints towards a closed universe! (In this case $\Omega _T$ is nearly
independent on $q_0$.) If we hope to obtain a flat universe, then we may
choose the point $P4=(0.36,0.64)$, which has $l_D=208c$ and $q_0=-0.10$ at
edge of observation values. It is notable that such as point can exist for
flat universe. The character of this point, as comparing with current favor
value, owns higher hubble constant, lower acceleration parameter, more 
right-side first peak (i.e.,
a little large $l_D$), higher cold dark mater density, lower vacuum energy
density, which is still flat universe. From this figure we can see if the
accuracy of $l_D$ is risen, how huge progress should be made for
establishment of $\Omega _T$ value. We shall wait for the exiting precious
results from the future Map and Planck.

We can see clearly the phenomenon claimed by Weinberg from this figure. The
first peak position $l_D$ determines the cosmic total density $\Omega _T$
sensitively. If the precision about $l_D$ of Boomerang measurement is
reliable, i.e., $190<l_D<202$ (see Ref.[1]), then we obtain a result $%
1.03<\Omega _T<1.08$ from this figure. This shows that our universe may be a
closed rather than flat$^{[9]}$! This will bring a great challenge for the
present elegant cosmological theories based on an eternal chaotic inflation$%
^{[10]}$. In another hand, if we review their conclusion $0.88<\Omega _T<1.12
$ (see Ref.[1] again) and suppose that all error of $l_D$ comes from
uncertainty of $\Omega _T$, we shall get an unexpected range $180<l_D<240$
from the just same figure. It is obvious that this error range of $l_D$ is
too large for current measurement. The Ref.[1] considered of course many
complicated factors (a little example is the error from $z_r$, $h$ and $%
\omega _b$), therefore in spit of the accuracy of $l_D$ is very high, we can
still only obtain the $\Omega _T$ value with very low accuracy.

\section{Conclusion}

Conforming to the Weinberg's thought, the formula (1) and its simplification
(Eqs.(6,8) and figure.1) supply for us a shortcut method to analysis error
origination to determine the cosmic total density. It can help us to
understand deeply the physical essential from the data of the CMBR
anisotropy. Weinberg's phenomenon, i.e, sensitivity of $\Omega _T$ with
respect to $l_D$, is very clear in our figure. In spit of our qualitative
analysis is available widely, however our concrete numerical result depends
seriously on the Efstathiou-Bond coefficient $C_{eb}$, we hope that people
can understand it deeply in the further investigation.

\newpage
\bigskip

{\bf Acknowledgment:}

{\hspace*{5mm} This work is supported by The foundation of National Nature
Science of China, No.19777103. The author would like to thank useful
discussions with Profs. J.-S.Chen, Z.-G.Deng, X.-M.Zhang Y.-Z.Zhang 
and Y.-Q.Yu. }

\bigskip

{\bf References:}

[1] P.de Bernardis et al., Nature 404(2000)955.

[2] M.Kamionkowski, D.N.Spergel and N.Sugiyama, Ap.J.426(1994)L57.

[3] N.A.Bahcall, J.P.Ostriler, S.Perlmutter and P.J.Steinhardt,

Science 28(1999)1481.

[4] S.Weinberg, astro-ph/0006276.

[5] C.-P.Ma and E.Bertschinger, Astrophys.J.455(1995)7.

[6] G.Efstathiou and J.R.Bond, astro-ph/9807103, in MNRAS.

[7] E.W. Kolb and M.S. Turner,

{\sl The Early Universe}, Addison Wesley, 1990. To see Eq.(3.89b).

[8] S.J.Perlmutter et al., Nature 391(1998)51;

A.G.Riess et al., Astron.J.116(1998)1009.

[9] M.White, astro-ph/0004385.

[10] A.H.Guth, astro-ph/0002156.

\bigskip

\begin{figure}[h]
{\epsfxsize=20cm\epsfysize=16.64cm \centerline{
\epsfbox{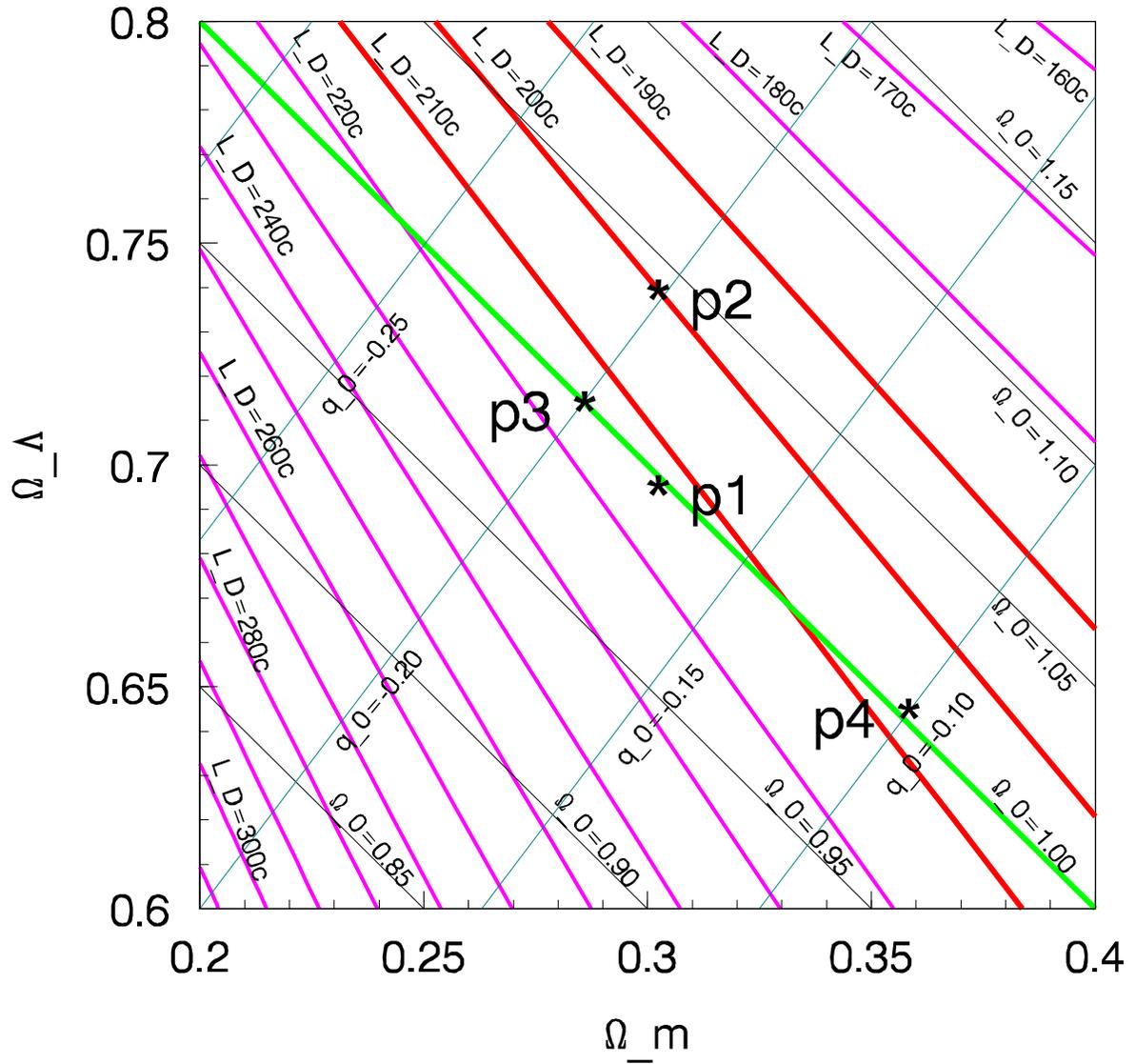}}}
\caption{\Large ~~~~~A contour graph of the first peak position $l_D$.}
\end{figure}

\end{document}